# THE GRAVITATIONAL BOHR RADIUS

Robert L. Oldershaw

Amherst College

Amherst, MA 01002 USA

ABSTRACT: The gravitational Bohr radius (GBR) characterizes the size of a hypothetical ground state hydrogen atom wherein the binding interaction between its nucleus and its electronic structure is purely gravitational. The conventional calculation of the GBR, based on the standard Newtonian gravitational coupling constant, yields an astronomical size for the "gravitational atom." On the other hand, a discrete fractal cosmological paradigm asserts that the gravitational coupling constant that applies *within* Atomic Scale systems is roughly 38 orders of magnitude larger than the conventional gravitational constant. According to calculations based on predictions of this discrete fractal paradigm, the value of the GBR is actually on the order of $2\pi a_o$, where $a_o$ is the standard Bohr radius. Implications of this revised gravitational Bohr radius are discussed.

1. INTRODUCTION

The Bohr radius ($a_0$), which is conventionally defined as the radius of maximum probability for finding the ground state electron in the hydrogen atom, is determined by the equation:

$$a_0 = 4\pi\varepsilon_0 \hbar^2/me^2 \,, \qquad (1)$$

where $\varepsilon_0$ is the electromagnetic permittivity constant, $\hbar$ is Planck's constant divided by $2\pi$, m is the mass of the electron, and e is the unit of electromagnetic charge for subatomic atomic particles. The value of $a_0$ is approximately $0.529 \times 10^{-8}$ cm. An interesting exercise in the physics of atomic systems is to calculate an approximate radius for the hydrogen atom in the hypothetical case that gravitational interactions, rather than purely electromagnetic interactions, govern the binding and dynamics of the atom. For this hypothetical case, we replace the electromagnetic terms $e^2/4\pi\varepsilon_0$ in the conventional calculation of $a_0$ with the equivalent gravitational terms GMm, where M is the mass of the proton. The gravitational Bohr radius (R) is then:

$$R \approx \hbar^2/GMm^2 \,, \qquad (2)$$

And the numerical value of R is $\approx 1.1 \times 10^{31}$ cm.

This enormous value for R is larger than the size of the observable universe, and is usually considered as solid proof that atoms are not governed by gravitational interactions. As we will see in this paper, however, the conventional calculation of R may involve a serious scaling error. A discrete fractal paradigm for the unbounded hierarchical organization of nature's systems proposes that gravitational interactions *within* Atomic Scale systems are roughly $10^{38}$ times stronger than the more familiar gravitational interactions within Stellar Scale systems. When the Atomic Scale gravitational constant predicted by the discrete fractal paradigm is used in Eq. (2), the corrected value for the gravitational Bohr radius is $\approx 2\pi a_0$, which is a very reasonable estimate for the outermost radial boundary of the ground state hydrogen atom.

## 2. DISCRETE FRACTAL SCALING FOR $G_\Psi$

The discrete fractal paradigm mentioned above is referred to as the Self-Similar Cosmological Paradigm[1] and it emphasizes nature's hierarchical organization of systems from the smallest observable subatomic particles to the largest superclusters of galaxies. The new fractal paradigm also highlights the fact that nature's global hierarchy is highly stratified into discrete Scales, of which we can currently observe the Atomic, Stellar and Galactic Scales. A third important principle of the fractal paradigm is that the cosmological Scales are rigorously self-similar, such that for each class of objects on a given Scale there is analogous class of objects on every other Scale. The self-similar analogues on different Scales have rigorously analogous morphologies, kinematics and dynamics. When the general self-similarity among the discrete Scales is exact, the paradigm is referred to as Discrete Scale Relativity[2] and nature's global spacetime geometry manifests a new universal symmetry principle: *discrete scale invariance*.

Based upon decades of studying the scaling relationships between analogues on the Atomic, Stellar and Galactic Scales, a close approximation to nature's actual Scale transformation equations for the length (L), time (T) and mass (M) parameters of analogue systems on neighboring Scales $\Psi$ and $\Psi$-1 are as follows.

$$L_\Psi = \Lambda L_{\Psi-1} \tag{3}$$

$$T_\Psi = \Lambda T_{\Psi-1} \tag{4}$$

$$M_\Psi = \Lambda^D M_{\Psi-1} \tag{5}$$

The self-similar scaling constants $\Lambda$ and D have been determined empirically[1] and are equal to $\cong$ 5.2 x $10^{17}$ and $\cong$ 3.174, respectively. Different cosmological Scales are designated by the discrete index $\Psi$ ($\equiv$ …, -2, -1, 0, 1, 2, …) and the Stellar Scale is usually assigned $\Psi=0$.

Since the discrete fractal scaling applies to *all dimensional parameters*, the Scale transformation equations also apply to dimensional "constants." Within the context of the discrete fractal paradigm, it has been shown[2] that the gravitational coupling constant $G_\Psi$ scales as follows.

$$G_\Psi = [\Lambda^{1-D}]^\Psi G_0 \tag{6}$$

Therefore the Atomic Scale value $G_{-1}$ is $\Lambda^{2.174}$ times $G_0$ and equals $\cong$ 2.18 x $10^{31}$ cm$^3$/g sec$^2$. When $G_{-1}$ is used in Eq. (2), instead of the Stellar Scale value of $G_0$, then,

$$R \cong 3.67 \times 10^{-8} \text{ cm} \approx 2\pi a_0. \tag{7}$$

3. IMPLICATIONS OF THE REVISED GRAVITATIONAL BOHR RADIUS

The wavefunction for the ground state hydrogen atom peaks at $a_0$, but tapers off to a larger effective limiting radius that has been estimated[1] to be on the order of $\approx 3a_0$. Therefore one might wonder about possible explanations for the apparent factor of $\approx 2$ difference between the calculated value of $\approx 2\pi a_0$ and the empirical value of $\approx 3a_0$. Two physically realistic potential explanations are listed below.

   a. Since electromagnetic interactions are ignored in calculating R, and since these additional interactions would increase the binding strength between the nuclear and the electronic matter, one would expect R to be an overestimate of the limiting radius of an actual hydrogen atom.

   b. The above calculation of R is based on a Newtonian gravitational model, which in most cases is only a first approximation to General Relativity. It is very reasonable to expect that the transition to a full general relativistic calculation might involve a correction factor of $\approx 2$.

The major implication of the GBR calculation using the gravitational coupling constant predicted by the discrete fractal paradigm is that atoms might be governed primarily, although not exclusively, by gravitational dynamics. The discrete fractal paradigm and the discrete cosmological scaling upon which the paradigm is based are backed up by a considerable amount of empirical evidence, including a long list of retrodictive successes[1] and some very encouraging tentative results pertaining to its major predictions[3,4] regarding the specific nature of the enigmatic dark matter that dominates the mass of the observable universe. The concept that gravitation is the primary dynamical interaction *within* bound Atomic Scale systems is also supported by two recent papers discussing the concept of Discrete Scale Relativity. One paper[5] demonstrates that hadrons can be modeled as Kerr-Newman black holes when $G_{-1}$ is adopted as the relevant gravitational coupling constant. The other paper[6] argues that the adoption of $G_{-1}$ for Atomic Scale systems leads to an explanation for the long-standing mystery surrounding the physical meaning of the fine structure constant. The latter is shown to be the ratio of the strengths of the unit electromagnetic and gravitational interactions *within bound* atomic systems. This implies that within such systems the gravitational interactions are about 137.036 times stronger than electromagnetic interactions. Since $G_0$ applies in the case of *external* interactions between *unbound* charged Atomic Scale systems, their dynamics are governed almost entirely by electromagnetic interactions *so long as they remain unbound*, as is observed.

REFERENCES

1. Oldershaw, R.L., *International Journal of Theoretical Physics*, **28**(6), 669-694 and **28**(12), 1503-1532, 1989; also see http://www.amherst.edu/~rloldershaw which is the most comprehensive resource for studying the SSCP.

2. Oldershaw, R.L., *Astrophysics and Space Science*, **311**(4), 431-433, 2007 [DOI: 10.107/s10509-007-9557-x]; also available at http://arxiv.org as arXiv:physics/0701132v3.

3. Oldershaw, R.L., *Astrophysical Journal*, **322**(1), 34-36, 1987.

4. Oldershaw, R.L., *Fractals*, **10**(1), 27-38, 2002.

5. Oldershaw, R.L., *Astrophysics and Space Science*, submitted, 2008; also available at http://arxiv.org as arXiv:astro-ph/0701006v2.

6. Oldershaw, R.L., *Electronic Journal of Theoretical Physics*, submitted, 2007; also available at http://arxiv.org as arXiv:0708.3501v1 [physics.gen-ph].